\documentclass[
superscriptaddress,
nobibnotes,
 amsmath,amssymb,
 aps,
prb,
twocolumn,
]{revtex4-1}

\usepackage{graphicx}
\usepackage{dcolumn}
\usepackage{bm}
\usepackage{lipsum}
\usepackage{siunitx}
\usepackage{hyperref}
\usepackage{braket}
\hypersetup{
    colorlinks=true,
    linkcolor=blue,
    filecolor=magenta,      
    urlcolor=blue,
}

\newcommand{\hc}[1]{{#1}^{\dagger}}

\begin{document}

\title{Effects of reduced dimensionality, crystal field, electron-lattice coupling, and strain on the ground-state of a rare-earth nickelates monolayer}

\affiliation{Stewart Blusson Quantum Matter Institute, University of
British Columbia, Vancouver BC V6T 1Z4, Canada}
\affiliation{Department of Physics, McGill University, Montréal, QC H3A 2T8, Canada}
\affiliation{Department of Physics and Astronomy, University of
British Columbia, Vancouver BC V6T 1Z1, Canada} 
\affiliation{Max Planck Institute for Solid State Research, Heisenbergstrasse 1, D-70569 Stuttgart, Germany}
\affiliation{These authors contributed equally to this work.}

\author{Rodrigo \surname{Chavez Zavaleta}}
\email{rodrigo.chavezzavaleta@mail.mcgill.ca}
\affiliation{Stewart Blusson Quantum Matter Institute, University of
British Columbia, Vancouver BC V6T 1Z4, Canada}
\affiliation{Department of Physics, McGill University, Montréal, QC H3A 2T8, Canada}
\affiliation{These authors contributed equally to this work.}

\author{Stepan Fomichev}
\email{fomichev@phas.ubc.ca}
\affiliation{Stewart Blusson Quantum Matter Institute, University of
British Columbia, Vancouver BC V6T 1Z4, Canada}
\affiliation{Department of Physics and Astronomy, University of
British Columbia, Vancouver BC V6T 1Z1, Canada} 
\affiliation{These authors contributed equally to this work.}

\author{Giniyat Khaliullin}
\affiliation{Max Planck Institute for Solid State Research, Heisenbergstrasse 1, D-70569 Stuttgart, Germany}

\author{Mona Berciu}
\affiliation{Stewart Blusson Quantum Matter Institute, University of
British Columbia, Vancouver BC V6T 1Z4, Canada}
\affiliation{Department of Physics and Astronomy, University of
British Columbia, Vancouver BC V6T 1Z1, Canada} 

\date{October 3, 2021}

\begin{abstract}
Motivated by the potential for cuprate-like superconductivity in monolayer  rare-earth nickelate superlattices, we study the effects of crystal field splitting, lattice distortions and strain on the charge, magnetic, and orbital order in undoped two-dimensional (2D) nickelate monolayers $R$NiO$_3$. We use a two-band Hubbard model to describe the low-energy electron states, with correlations controlled by a effective Hubbard $U$ and Hund's $J$. The electrons are coupled to the octahedral breathing-mode lattice distortions. Treating the lattice semiclassically, we apply the Hartree-Fock approximation to obtain the phase diagram for the ground-state as a function of the various parameters. We find that the 2D confinement leads to strong preference for the planar $d_{x^2-y^2}$ orbital even in the absence of a crystal-field splitting. The $d_{x^2-y^2}$ polarization is enhanced by adding a crystal field splitting, whereas coupling to breathing-mode lattice distortions weakens it. However, the former effect is stronger, leading to $d_{x^2-y^2}$ orbital and antiferromagnetic (AFM) order at reasonable values of $U,J$ and thus to the possibility to realize cuprate-like superconductivity in this 2D material upon doping.
We also find that the application of tensile strain enhances the cuprate-like phase and phases with orbital polarization in general, by reducing the $t_2 / t_1$ ratio of next-nearest to nearest neighbour hopping. On the contrary, systems with compressive stress have an increased hopping ratio and consequently show a preference for ferromagnetic (FM) phases, including, unexpectedly, the out-of-plane $d_{3z^2-r^2}$ FM phase.
\end{abstract}

\maketitle

\section{Introduction}\label{sec:intro}

The origin of the high-temperature superconductivity in the cuprate family remains an open question, despite over three decades of intense efforts to find the answer.\cite{Anderson_2007,Chu_2015} Given the inherent difficulties both in identifying the proper minimal model for such complex materials, and in dealing with the resulting strongly correlated problem, it has long been recognized that it would be valuable to have access to other classes of materials that contain similar "building blocks" with those believed to be key for cuprate physics. As a minimal starting point, this would require  layers hosting a square lattice of spins-${1\over 2}$ that are antiferromagnetically coupled in the stoichiometric (undoped) material, and the ability to dope them.
  
In this context, given the proximity  of Ni and Cu in the table of elements, Anisimov \textit{et al.}\cite{Anisimov_1999} investigated the possibility of finding nickelate analogs to the cuprates. They concluded that if the Ni$^+$ oxidation state can be stabilized into the appropriate planar coordination with oxygen, as would be expected in \textit{infinite layer} compounds such as LaNiO$_2$, the electronic structure of the undoped NiO$_2$ layers should be similar to that of CuO$_2$ layers, and presumably superconductivity would appear upon hole-doping. This prediction was confirmed very recently, when it was revealed that NdNiO$_2$ indeed becomes superconducting upon doping, albeit with a rather moderate $T_c \sim \SI{15}{\kelvin}$.\cite{Li_2019,Sawatzky_2019,Zeng_2020} Given the identical $3d^9$ electronic configurations of the Ni$^+$ and Cu$^{2+}$ oxidation states in the corresponding undoped compounds, at first this may appear as a foregone conclusion. However, the bigger charge-transfer energy in the $d^9$ nickelates should lower the degree of hybridization between $3d$ and $2p$ orbitals, suggesting that Ni $3d^9$- O $2p$ bonds are not as strongly covalent as those in the CuO$_2$ layer. Because of this and other complications related to the existence of Nd bands crossing the Fermi energy,\cite{Lee_2004} the mechanism of this new superconducting state is likely to remain controversial for some time \cite{Hepting_2020,Zhang_2020a,Zhang_2020b,Wu_2020,Leonov_2020,Karp_2020,Jiang_2020,Hirsch_2019}. 

The other scenario that might lead to a nickelate analog for cuprates is based on the Ni$^{3+} (3d^7)$ oxidation state.\cite{Chaloupka_2008,Hansmann_2009} Unlike the cuprates, it has one electron instead of one hole in the $e_g$ manifold, but the two situations are formally equivalent if $e_g$ degeneracy is lifted and the $d_{x^2-y^2}$ orbital becomes half-filled. The increased oxidation leads to a significantly decreased charge-transfer energy, and thus a much higher degree of covalency between the $3d_{x^2-y^2}$ and the $2p$ orbitals than in the infinite-layer $d^9$ nickelates -- and likely also more than in the cuprates.

On one hand, a large $pd$-covalency in $d^7$ nickelates is a positive
factor in favor of Zhang-Rice singlet physics, believed to be essential in cuprates. However,
it also supports charge-ordering tendencies, which are well observed in
the rare-earth nickelates $R$NiO$_3$. In these three-dimensional perovskites,
wide $pd\sigma$ bands of $e_g$ symmetry are formed, containing finite
density of holes on the oxygen sites which strongly couple to Ni-spins
to form the Zhang-Rice singlets. At low temperatures, charge order
sets in, which can be viewed as a condensation of ``self-doped''
Zhang-Rice singlets on every second Ni ion; plus, the order is strongly
supported by breathing-type lattice distortion. This discourages
Jahn-Teller physics and related orbital order in three-dimensional $d^7$
nickelates.

In cuprates, a large orbital polarization in favor of a planar
$d_{x^2-y^2}$ state is due to their layered structure, with quasi-two
dimensional hopping geometry and large $e_g$ orbital splitting in such
lattices. Along these lines, it was proposed to achieve a single-band of
$d_{x^2-y^2}$ symmetry in $d^7$ nickelate heterostructures. Such heterostructures would consist of LaNiO$_3$ monolayers, separated from each other by insulating block layers such as SrTiO$_3$ or LaAlO$_3$.\cite{Chaloupka_2008,Hansmann_2009} This would inhibit $c$-axis hopping, thus discouraging the occupancy of the $d_{3z^2-r^2}$ orbital; meanwhile tensile strain from lattice mismatch would compress the NiO$_2$ octahedra  along the $c$ axis and lower the energy of the $d_{x^2-y^2}$ orbital, which would become half-filled like in the undoped cuprates. Subsequent DFT+DMFT calculations predicted orbital polarization as large as 30\% in such structures, provided the insulating block layer is tuned appropriately \cite{Peil_2014,Han_2010}. 

It is important to note that to date, such single-layer, disorder-free superlattices have proved difficult to grow, so their magnetic, charge and lattice properties are still unknown. To the best of our knowledge, the only such heterostructures have been reported in Ref. \onlinecite{Disa_2017}, with monolayers of NdNiO$_3$ embedded in a matrix of NdAlO$_3$. The authors found very different behaviour of these heterostructures when compared to even those with double layers of  NdNiO$_3$. In particular, the ground state was found to be insulating and without signs of either magnetic or charge order, whereas double-layer and thicker layer samples showed both antiferromagnetic and charge order reminiscent of the bulk properties of NdNiO$_3$. In these samples, the strain is compressive and a negative crystal field splitting favors the occupation of the $3d_{3z^2-r^2}$ orbital, so they are not a good candidate for a cuprate analog. Nevertheless, the fact that the monolayer physics is so different from that of multi-layer/bulk  samples suggests that similar findings might be expected in heterostructures more suitable to produce cuprate analogs. In fact, 
numerous \textit{multi-layer} LaNiO$_3$ heterostructures have been successfully synthesized, and were found to have orbital polarizations  ranging from 0  up to $\sim 50\%$ \cite{Freeland_2011, Benckiser_2011, Wu_2013, Disa_2015} depending on the specifics of interlayer spacers and tensile/compressive stress.

Another important ingredient in these materials is the presence of strong electron-lattice coupling. Its role for the rare-earth nickelates is well established:\cite{Johnston_2014,Bisogni_2016,Mizokawa_2000,Park_2012,Lau_2013,Puggioni_2012,Caviglia_2012,Catalano_2018,Fomichev_2020} it is responsible for the breathing-mode alternation of compressed and expanded oxygen octahedra in the ground state of (most) bulk nickelates. Whether such distortions are present in the ground-state of the proposed 2D monolayer heterostructure is an open question, and an important one given that this might interfere with cuprate-like physics.

There are, however, clues that lattice coupling does play a big role in the multi-layer heterostructures. One such clue is the peculiar effect of strain on orbital polarization and occupancy. The leading effect of applied volume-preserving strain is that it can tune the energies of the various orbitals by altering the crystal field symmetry (the so-called strain-induced orbital polarization, SIOP model \cite{Middey_2016}) -- in the case of nickelates, the symmetry of the oxygen octahedral cage around the nickel atom. In particular, $xy$ plane compressive strain should inhibit the in-plane orbital in favour of the out-of-plane $d_{3z^2-r^2}$, and tensile $xy$ plane strain should do the opposite. While ultrathin ($\sim$ 10 unit cells) films of LaNiO$_3$ under compressive strain (on LaAlO$_3$ substrates) indeed show a lowering of $d_{3z^2-r^2}$ energy \cite{Chakhalian_2011}, tensile strain in the same material (from SrTiO$_3$ substrates) yields the breathing-mode distortion and little-to-no orbital  polarization. Similarly, in LaNiO$_3$ / LaAlO$_3$ superlattices, compressive strain gives rise to the peculiar behaviour where the $e_g$ doublet has an energy splitting on the order of \SI{100}{\meV}, but no orbital polarization has been detected \cite{Freeland_2011}. Meanwhile tensile strain gives the opposite -- equal energies, but a mild (5\%) polarization were observed. All this hints at strong lattice involvement.

In this paper we investigate the effects of reduced dimensionality, crystal field, strain and electron-lattice coupling on the ground state of a  nickelate monolayer, to see in what circumstances these may favor a cuprate-like ground-state. Following Lee \textit{et al.}\cite{Lee_2011} and Subedi \textit{et al.} \cite{Subedi_2015}, we adopt a phenomenological two-band ($e_g$) Hubbard-Kanamori model to describe the electronic structure of the monolayer. We also include the coupling to breathing-mode distortions, which are treated semiclassically and coupled to electrons via a Holstein-like term.\cite{Fomichev_2020} Using this model, we show that the reduction from 3D to 2D promotes orbital occupancy of the $d_{x^2-y^2}$ orbitals, as they lower their energy through in-plane hopping relative to the $d_{3z^2-r^2}$ which cannot hop as freely. This effect alone, even in the absence of favorable crystal-field splitting, is enough to give a cuprate-like phase, ensuring the preferential occupancy of $d_{x^2-y^2}$ and AFM order for reasonable values of Hubbard repulsion $U$ and Hund's coupling $J$. This cuprate-like phase could not be stabilized in 3D,\cite{Fomichev_2020} but appears naturally in nickelate monolayers.

We then analyze the interplay between crystal field splitting (which favors orbital polarization), electron-lattice coupling (which favors ``cuprate-competitor'' phases that exhibit breathing-mode distortion) and strain, on the stability of this cuprate-like phase. Our findings indicate that the negative influence of the electron-phonon coupling is relatively weak, and can be easily compensated by the crystal-field splittings expected in such heterostructures. The role of strain is quite intricate, but in qualitative agreement with the surprising experimental findings in the multi-layer heterostructures, discussed above.

The remainder of this paper is structured as follows. In section \ref{sec:model}, we introduce our Hamiltonian and comment on its suitability to describe these complex materials. In section \ref{sec:calc}, we briefly describe the mean-field approximation we use to study this model. Our results are presented in Sec. \ref{sec:results}, and the conclusions and outlook are discussed in Sec. \ref{sec:concl}. The Appendices contain additional information about our model and its numerical solution. 

\section{Model}\label{sec:model}
We consider a 2D square lattice with the lattice constant $a$ set to 1. We define the Hamiltonian  as follows:
\begin{equation}
\hat H = \hat T + \hat H_{e-e} + \hat H_{e-l} +  \hat H_{l},
\label{eq:basic}
\end{equation}
which comprises the band energy of $e_g$ electrons $\hat{T}$, their mutual interactions $\hat{H}_{e-e}$, lattice energy $H_l$, and the electron-lattice coupling $\hat{H}_{e-l}$. Like in the bulk case \cite{Lee_2011,Subedi_2015,Fomichev_2020}, the two active orbitals per site are taken to have the symmetries of the Ni $e_g$ doublet; we discuss this more at the end of this section, where we comment on the applicability of this simple model to nickelates. We focus on a two-site unit cell that allows for (simple) magnetic and orbital order: the unit cell is coordinated through the rocksalt-type wavevector $\mathbf{Q}_c = \pi(1,1)$ to represent the symmetry of the charge order and of the breathing-mode distortion experimentally observed in ultrathin films and superlattices \cite{Fursich_2019}. We note that a generalization to a 4-site unit cell is possible and was used in Ref. \onlinecite{Fomichev_2020} for the study of 4-site magnetic order in 3D bulk systems. However, here our main interest is the appearance of cuprate-like physics, and this can be addressed using a 2-site unit cell.

For the two orbitals we introduce the shorthand notation $\ket{z} = \ket{3z^2 - r^2}, \ket{\bar{z}} = \ket{x^2 - y^2}$, with the associated  operators $\hc{d}_{ia\sigma}$ creating an electron at site $i$ in the orbital $a = z, \bar{z}$ with spin $\sigma$. All hopping in the lattice then proceeds through these orbitals, as shown elsewhere \cite{Fomichev_2019}: we include hopping for the first, second, and third nearest neighbours, with corresponding ``un-strained'' hopping integrals $t_1^{(0)}, t_2^{(0)}, t_3^{(0)}$ restricted to the 2D plane. (Note that third nearest neighbour hopping on a 2D lattice corresponds to fourth nearest neighbour coupling on a 3D lattice. This accounts for some differences between the work here and the 3D expressions in Ref. \onlinecite{Fomichev_2020}). The energy difference between the active orbitals, due to Jahn-Teller $c$ axis symmetry breaking, is modelled through a crystal field splitting term $\Delta_{\text{CF}}$. The non-interacting part of the electronic Hamiltonian is thus:
\begin{multline}
            \hat T = -\sum_{\langle ij \rangle ab\sigma} t_{ia,jb}(d_{ia \sigma}^{\dagger} d_{jb\sigma} + \text{H.c.}) + \\ 
            + \frac{\Delta_{\text{CF}}}{2} \sum_{i\sigma} (d_{iz \sigma}^{\dagger} d_{iz\sigma} - d_{i \bar z \sigma}^{\dagger} d_{i \bar z \sigma}).
\end{multline}

Making use of crystal symmetry together with periodic boundary conditions, this tight-binding contribution can be Fourier transformed according to 
\begin{equation}
    \hc{d}_{\mathbf{k}+\zeta \mathbf{Q}_c, a\sigma} = \frac{1}{\sqrt{N}} \sum_i e^{i (\mathbf{k} +\zeta \mathbf{Q}_c) \mathbf{R}_i} \hc{d}_{ia\sigma}
\end{equation}
where $N$ is the number of unit cells, $i$ is the lattice site index, and the momentum  $\mathbf{k}$ is inside the diamond-shaped Brillouin zone with corners at $(0, \pm \pi  ), (\pm \pi, 0)$. The $\mathbf{Q}_c = (\pi,\pi)$ wavevector (together with $\zeta = 0,1$) serves to distinguish two types of momentum operators obtained by folding the square Brillouin zone into the diamond shape. This Brillouin zone is chosen due to the breathing-mode symmetry of both charge order and lattice distortions in the nickelates, as given by the ordering wavevector $\mathbf{Q}_c$ found experimentally. As shown elsewhere \cite{Fomichev_2019}, the tight-binding part of the Hamiltonian then may be written as
\begin{multline}
            \hat T = \sum_{\mathbf{k} \zeta ab} t_{ab}(\mathbf{k}+\zeta\mathbf{Q}_c) \hc{d}_{\mathbf{k} + \zeta, a\sigma} d_{\mathbf{k}+\zeta, b \sigma} + \\ 
            + \frac{\Delta_{\text{CF}}}{2} \sum_{\mathbf{k} \zeta\sigma} (d_{\mathbf{k}+\zeta,z \sigma}^{\dagger} d_{\mathbf{k}+\zeta,z\sigma} - d_{\mathbf{k}+\zeta, \bar z \sigma}^{\dagger} d_{\mathbf{k}+\zeta \bar z \sigma}).
\end{multline}
where the $t_{ab}(\mathbf{k})$ expressions are listed in Appendix \ref{app:Hopping}.

A key feature of thin nickelate films grown on different substrates is the strain induced in the film by the lattice constant mismatch. A standard approach to incorporate strain into a model of this type is through a parametrization of the hopping integrals (see, for example, Pereira \textit{et al} in Ref. \onlinecite{Pereira_2009}). We choose to model the effects of biaxial physical strain through exponential decay
\begin{equation}
    t_i = t_i^{(0)} \cdot e^{- \epsilon \tau \lambda_i},
    \label{eq:strain}
\end{equation}
where $\epsilon$ is the strain. On geometric grounds, the distance factor $\lambda$ is 1 for the nearest neighbour, $\sqrt{2}$ for the second nearest neighbour, and 2 for the third nearest neighbour. We set the free parameter $\tau$, which controls the rate of hopping integral decay with strain, to $1$ and treat strain $\epsilon$ as a free (dimensionless) parameter. 

To model the electron-electron interactions, we use the two-band Hubbard-Kanamori Hamiltonian:
\begin{multline}
    \hat H_{e-e} = U \sum_{ia} \hat n_{ia\uparrow}\hat n_{ia\downarrow} + (U- 2J)\sum_{i, \sigma} \hat n_{iz\sigma}\hat n_{i \bar{z}\bar{\sigma}} + \\
    + (U - 3J)\sum_{i\sigma} \hat n_{iz\sigma}\hat n_{i \bar{z}\sigma} - J\sum_{i \sigma}d_{iz \sigma}^{\dagger} d_{iz\bar{\sigma}}d_{i\bar{z}\bar{\sigma}}^{\dagger} d_{i\bar{z}\sigma} + \\
    + J\sum_{i a}d_{ia \uparrow}^{\dagger} d_{ia\uparrow}d_{i\bar{a}\downarrow}^{\dagger} d_{i\bar{a}\downarrow}.
    \end{multline}
Here $\hat{n}_{ia\sigma}$ is the electron number operator, measuring the electron density at site $i$, in orbital $a$ with spin $\sigma$.

The electron-lattice on-site interaction is described using the Holstein model
\begin{equation}
    \hat{H}_{e-l} = -2\epsilon_b \sum_{j}u_j\left(\sum_{a\sigma}\hat{n}_{ja\sigma}-1\right)
\end{equation}
The dimensionless $u_j= U_j k/g$ characterizes the isotropic distortion $U_j$ of the octahedral oxygen cage around the Ni site $j$,  treated semi-classically\cite{Fomichev_2020}, when the electron number differs from the average electron number of 1. (This corresponds to a quarter-filled $e_g$ doublet, \textit{i.e.} the $3d^7$ configuration.) Here, $k$ is the lattice spring constant (see below), $g$ is the Holstein coupling, and $\epsilon_b=g^2/(2k)$.

Finally, the lattice energy in the semi-classical approximation is:\cite{Fomichev_2020}

\begin{equation}
    H_{l} = \epsilon_b \sum_j \left( u_j^2 + \frac{\alpha}{2} u_j^4 \right).
\end{equation}
where $\alpha$ represents lattice anharmonicity. This form is obtained from the standard semiclassical lattice energy $H_{l} = \sum_j \left( \frac{k}{2} U_j^2 + \frac{A}{4} U_j^4 \right)$ upon rewriting it in terms of the dimensionless $u_j= U_j k/g$. Our results are in the $|u|\ll 1$ limit, so the anharmonicity parameter $\alpha\propto A$ has little effect. For simplicity, we set it to $\alpha = 1$  in what follows.

A natural question is whether this relatively simple two-band model captures the complex electron behaviour in nickelates. Specifically, there is a growing consensus that the nickelates are a negative charge-transfer (NCT) material in the Zaanen-Sawatzky-Allen classification scheme \cite{Johnston_2014,Bisogni_2016,Mizokawa_2000,Park_2012,Lau_2013,Puggioni_2012,Caviglia_2012}. Instead of full O $2p^6$ bands and Ni in the $3d^7$ configuration (or perhaps a charge-disproportionated option\cite{Catalano_2018,Catalan_2008,Medarde_2009} where $3d^7 \, 3d^7 \rightarrow 3d^{7+\delta} \, 3d^{7-\delta}$), the relevant electronic structure is $3d^{8} \underline{L}$, with a ligand hole $\underline{L}$ on the oxygens for each Ni site. This scenario is supported by some \textit{ab initio} calculations \cite{Johnston_2014,Han_2011,Park_2012,Lau_2013} as well as experimental findings \cite{Bisogni_2016}, and seems to present a difficulty for our two-band model, which does not explicitly include O $2p$ orbitals.

However, Lee {\it et al.}\cite{Lee_2011} and Subedi {\it et al.}\cite{Subedi_2015} have shown that such a phenomenological two-band description of the nickelates is indeed reasonable, and gives  good agreement with available angle-resolved photoemission spectroscopy measurements.\cite{Eguchi_2009,Uchida_2011} The molecular orbitals of type $3d^{8} \underline{L}$ (formed through strong Ni-O hybridization) adopt the $e_g$-like symmetries because such symmetries maximize $pd$ overlap and allow holes to move into the oxygen $p$ band. Thus, in our model the two $e_g$ orbitals should be identified with these Ni-O molecular states. Because of this identification, the interactions $U, J$ do not take atomic-like values: they are instead strongly renormalized, and we treat them as free parameters (for a more detailed discussion, see Ref. \onlinecite{Fomichev_2020}).

\section{Hartree-Fock approximation}\label{sec:calc}

We treat our Hamiltonian in Eq. \ref{eq:basic} in the mean-field approximation. We are interested in the $T=0$ phase diagram, therefore we minimize the Hartree-Fock energy $E_{HF} = \bra{\Psi_e} \hat{H}(u_j) \ket{\Psi_e}$ (hereafter we define $\langle \hat{O} \rangle_e = \bra{\Psi_e} \hat{O} \ket{\Psi_e}$ for brevity, where $\hat{O}$ is any electronic operator) by choosing the most optimal Slater determinant $\ket{\Psi_e}$ to approximate the electron wavefunction, and $u_j$'s to describe the lattice distortions. We now briefly review the relevant details.

\subsection{Lattice contributions}

Due to the semiclassical treatment of the lattice, it is easiest to start by minimizing the energy with respect to the lattice distortions $u_j$. As shown in detail elsewhere \cite{Fomichev_2020}, this can be done with the help of Hellmann-Feynman theorem to yield a self-consistency equation
\begin{equation}\label{eq:latt-breath}
u_j + u_j^3 = \langle \hat{n}_j \rangle_e - 1. 
\end{equation}
The dominant lattice mode is the breathing-mode distortion, described by $u_j = u e^{i \mathbf{Q}_c \cdot \mathbf{R}_i}$, with the ordering wavevector in 2D being $\mathbf{Q}_c = \pi(1,1)$. Charge order is expected to have the same breathing-mode symmetry $\langle \hat{n}_j \rangle = 1 + \delta e^{i \mathbf{Q}_c \cdot \mathbf{R}_i}$, leading to
\begin{equation}
u + u^3 = \delta.
\end{equation}
This equation links the presence of extra electronic charge on a site and that site's lattice distortion. Its exact solution
\begin{equation}
u = \delta \frac{3}{2 \beta^{\frac{1}{3}}} \left[ \left( \sqrt{1 + \frac{1}{\beta}} + 1 \right)^{\frac{1}{3}} - \left( \sqrt{1 + \frac{1}{\beta}} - 1 \right)^{\frac{1}{3}} \right]
\end{equation}
with $\beta = \frac{27}{4} \delta^2$, allows us to eliminate the lattice degrees of freedom from the problem entirely, by expressing them in terms of the electronic charge modulation $\delta$.

Incidentally, we note that Jahn-Teller (JT) phonon modes could also be activated by the presence of the interlayers in the heterostructure (weak apical oxygen). If treated semiclassically, the impact of a JT lattice distortion would mostly renormalize the explicit crystal field term $\Delta_{\text{CF}}$, as it would produce an equation similar to Eq. \ref{eq:latt-breath}, with orbital polarization $O_{\text{FO}}$ playing the role of charge disproportionation $\delta$. As such, we do not explicitly include the JT mode, and instead vary the value of $\Delta_{\text{CF}}$ to represent various strengths of the Jahn-Teller coupling. 

\subsection{Electronic contributions}

The Hartree-Fock Slater determinant has the form
\begin{equation}
    \ket{\Psi_e} = \prod_n \hc{c}_n \ket{0},
\end{equation}
and is built up of \textit{a priori} unknown occupied electron orbitals $n$ with electron creation operator $\hc{c}_n$, in terms of which the effective Hartree-Fock Hamiltonian is diagonal. A unitary transformation connects the $\hc{c}_n$ orbitals to the original orbitals $\hc{d}_{ia\sigma}$ identified in Sec. \ref{sec:model}, by
\begin{equation}
    \hc{d}_{ia\sigma} = \sum_n \phi_n^*(ia\sigma) \hc{c}_n.
\end{equation}
By minimizing the total energy $E_{\text{HF}} = \langle \hat{H} \rangle_e$, we obtain a set of self-consistency equations for the unknown coefficients $\phi_n(ia\sigma)$ of the unitary transformation. More specifically, when the expectation value is evaluated by means of Wick's theorem, various mean field parameters $\langle \hc{d}_{ia\sigma} d_{ib\tau} \rangle_e$  arise (only on-site terms are non-zero because only local electron-electron and electron-lattice interactions are included in the model). They correspond to physical observables such as the on-site charge density $\langle \hat{n}_j \rangle_e = \sum_{a\sigma} \langle \hc{d}_{ja\sigma} d_{ja\sigma} \rangle_e = \sum_{n a\sigma} \phi^*_n(ja\sigma) \phi_n(ja\sigma)$, where the latter sum includes only the states $n$ occupied at $T=0$.

In this study we choose a 2-site unit cell coordinated in a breathing-mode manner, so we constrain the values of the mean-field parameters as follows

\begin{widetext}
\begin{align}
\langle \hc{d}_{ia\sigma} d_{ia\sigma} \rangle_e = &\frac{1}{4} \left[ 1 + \delta e^{i\mathbf{Q}_c\cdot \mathbf{R}_i}  \right] + \frac{a}{2} \Big[ O_{\text{FO}} + O_{\text{AF}} e^{i \mathbf{Q}_c \cdot \mathbf{R}_i} \Big] + \frac{\sigma}{2} \Big[ S_{\text{FM}} + S_{\text{AF}}e^{i\mathbf{Q}_c\cdot \mathbf{R}_i}\Big], \label{eq:mfp1}\\
\langle \hc{d}_{ia\sigma} d_{ia\bar{\sigma}} \rangle_e =& \langle \hc{d}_{ia\sigma} d_{i\bar{a}\sigma} \rangle_e = \langle \hc{d}_{ia\sigma} d_{i\bar{a}\bar{\sigma}} \rangle_e = 0. \label{eq:mfp4}
\end{align}
\end{widetext}
We neglected a large number of potential mean fields lying in the off-diagonal spin-orbital sectors, as experimental evidence indicates that orbital order beyond simple polarization is unlikely. 

In contrast with our previous 3D study \cite{Fomichev_2020}, explicit orbital polarization $O_{\text{FO}}$ (and its breathing-mode-like modulations, $O_{\text{AF}}$) are now allowed. This is because whereas orbital polarization has not been observed in bulk nickelates, thin-film nickelates and superlattices routinely display orbital energy shifts and occupancy differences. 

Equations (\ref{eq:mfp1})-(\ref{eq:mfp4}) represent a strong constraint on the possible spin, charge, orbital, and magnetic orders, while at the same time wrapping the difficult-to-parse matrix elements like $\langle \hc{d}_{ia\sigma} d_{ib\tau} \rangle_e$ into physical observables with readily understood meaning. For example, $O_{\text{FO}}$ and $O_{\text{AF}}$ together describe orbital polarization:
\begin{multline}
P_i = \sum_{\sigma} \left( \langle \hc{d}_{iz\sigma} d_{iz\sigma} \rangle_e - \langle \hc{d}_{i\bar{z}\sigma} d_{i\bar{z}\sigma} \rangle_e \right) = \\ = O_{\text{FO}} + O_{\text{AF}} e^{i \mathbf{Q}_c \cdot \mathbf{R}_i}.
\end{multline}

Using these prescriptions for $\langle \hc{d}_{ia\sigma} d_{ib\tau} \rangle_e$, and minimizing the total energy with respect to $\phi_n(ia\sigma)$, we derive the Hartree-Fock equations, which will depend on the mean field parameters $\delta, S_{\text{FM}}, O_{\text{FO}}$, etc. These equations can be equivalently seen as arising from a non-interacting Hartree-Fock Hamiltonian
\begin{widetext}
	\begin{eqnarray}\label{eq:hf-ham}
H_{\text{eff}} = &&\sum_{\mathbf{k}ab\sigma} t_{ab}(\mathbf{k})
\hc{c}_{\mathbf{k}a\sigma} c_{\mathbf{k}b\sigma} + \sum_{\mathbf{k}a\sigma} \Big[
  \frac{3U-5J}{4} - \frac{\sigma}{2} (U+J) S_{\text{FM}} + a \frac{-U + 5J}{2} O_{\text{FO}} + a\frac{\Delta_{\text{CF}}}{2} \Big]
\hc{c}_{\mathbf{k}a\sigma} c_{\mathbf{k}a\sigma} \nonumber\\ &&+
\sum_{\mathbf{k}a\sigma} \Big[ \frac{3U-5J}{4}\delta - 2\epsilon_b u - \frac{\sigma}{2} (U+J)
  S_{\text{AF}} + a \frac{-U + 5J}{2} O_{\text{AF}} \Big] \hc{c}_{\mathbf{k}+\mathbf{Q}_c,a\sigma}
c_{\mathbf{k}a\sigma}.
\end{eqnarray}
\end{widetext}
By defining $
\hc{\psi}_{\mathbf{k}}=
(\hc{\psi}_{\mathbf{k}z\uparrow},\hc{\psi}_{\mathbf{k}\bar{z}\uparrow},
\hc{\psi}_{\mathbf{k}z\downarrow}, \hc{\psi}_{\mathbf{k}\bar{z}\downarrow})$ and
$\hc{\psi}_{\mathbf{k}a\sigma} = (\hc{c}_{\mathbf{k}a\sigma},
\hc{c}_{\mathbf{k}+\mathbf{Q}_c,a\sigma})$, we  represent the Hamiltonian in Eq. (\ref{eq:hf-ham}) as a matrix $ H_{\text{eff}} =
\sum_{\mathbf{k}}^{} \hc{\psi}_{\mathbf{k}} h(\mathbf{k}) \psi_{\mathbf{k}}$ whose eigenvalues we seek. Note that this Hamiltonian is already block-diagonal, with the spin sectors completely decoupled, while the orbital sectors are only connected by the hopping matrix elements $t_{z\bar{z}}(\mathbf{k})$.

We solve these equations using an iterative approach: we make an initial guess for the mean field parameters $\mathbf{w}^{(0)} = (\delta, S_{\text{FM}}, O_{\text{FO}}, ...)$, solve the Hartree-Fock equations, construct the ground state and compute the resulting mean field parameters $\mathbf{v}^{(0)}$. The simplest choice is to then set $\mathbf{w}^{(1)}=\mathbf{v}^{(0)}$ and to iterate until convergence is achieved, {\it i.e.} the residual $\epsilon^{(n)} = |\mathbf{w}^{(n)} - \mathbf{v}^{(n)}|$ becomes smaller than a specified tolerance $\epsilon_0$. For mean-field problems such as this, with many mean-field parameters, this simplest choice $\mathbf{w}^{(n+1)}=\mathbf{v}^{(n)}$ turns out to be very poor. In Appendix \ref{sec:numerics} we present a very efficient choice, inspired by machine learning optimization, together with other salient computational details. 

\begin{figure}[t]
  \includegraphics[width=0.9\linewidth]{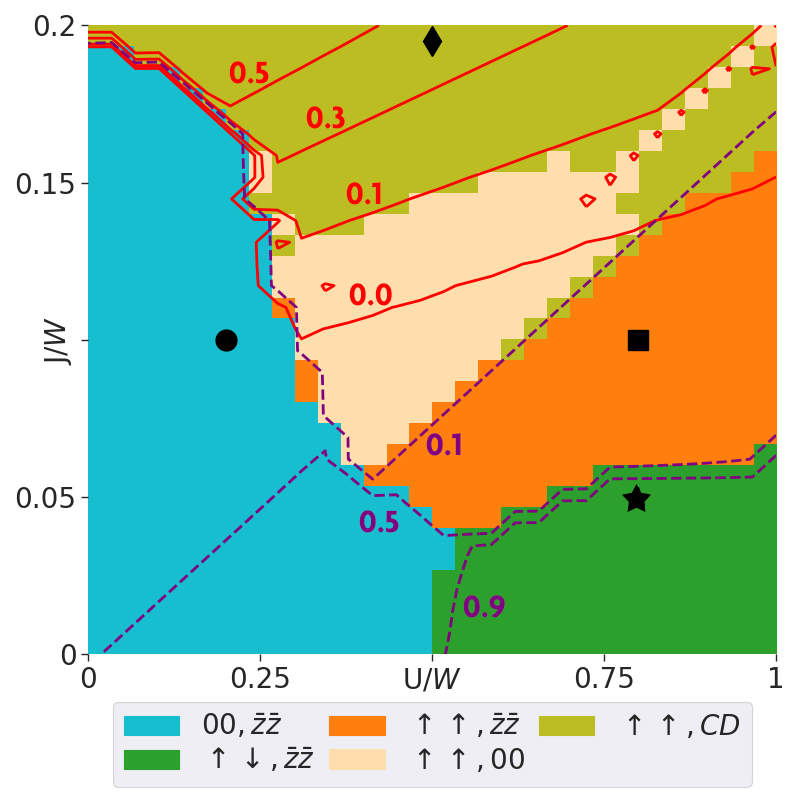}
  \caption{$T=0$ phase diagram of the 2D layer when $\epsilon_b=\Delta_{\text{CF}}=0$. $U$ and $J$ are scaled by $W$,  the bandwidth of the  non-interacting system.  Solid (red) contours show the CD $\delta$. Dashed (purple) contours show the orbital polarization $O_{\text{FO}}$. Resolution is 30 $\times$ 30. The four black symbols indicate the cases analyzed in more detail in Figs. (\ref{fig:cf_bulk_DOS1})-(\ref{fig:cf_bulk_DOS4}).}
  \label{fig:cf_bulk}
\end{figure}

\begin{figure}[b]
  \includegraphics[width=0.85\linewidth]{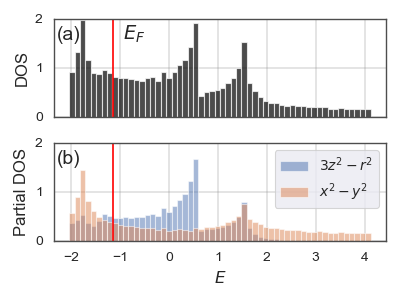}
  \caption{(a) Total density of states (DOS) per Ni site (in units of $1/t_1^{(0)}$), plotted against energy $E$ (in units of $t_1^{(0)}$), and (b) partial DOS per Ni site, projected onto the two orbitals, for the paramagnetic metallic phase $00,\bar{z}\bar{z}$ at $U/W = 0.2, J/W = 0.1$ (marked with a solid black circle in  Fig. \ref{fig:cf_bulk}).  The vertical red line marks the Fermi energy. In panel (b), the histograms' colors are labeled in the legend. Where they overlap, their transparency allows both histograms to be seen.}
  \label{fig:cf_bulk_DOS1}
\end{figure}

One final step is required, because the Hartree-Fock equations usually have multiple self-consistent solutions for the same set of parameters. We evaluate the total energy $E_{\text{HF}}$ for all these self-consistent solutions and choose the one with lowest energy as the best approximation for the ground state. For our model:
\begin{equation}
\label{eq:hf-energy}
\frac{E_{HF}}{N} = \frac{1}{N} \sum_{\text{occup} p} E_p - \frac{1}{N} \langle \hat{H}_{e-e} \rangle + \epsilon_b
\left(u^2 + \frac{u^4}{2} \right),
\end{equation}
where $E_p$ are the occupied eigenvalues of the Hartree-Fock Hamiltonian (\ref{eq:hf-ham}), and the Hubbard-Kanamori part of the energy is
\begin{multline}
\frac{1}{N} \langle \hat{H}_{e-e} \rangle =  \frac{3U-5J}{8} (1+\delta^2) - \frac{U+J}{2} \Big(
S_{\text{FM}}^2 + S_{\text{AF}}^2 \Big) + \\
+\frac{(-U+5J)}{2} \Big( O_{\text{FO}}^2 + O_{\text{AF}}^2 \Big).
\end{multline}

\section{Results}\label{sec:results}

\begin{figure}[b]
  \includegraphics[width=0.85\linewidth]{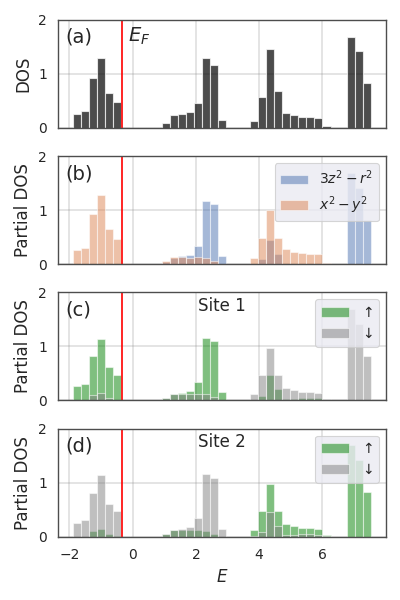}
  \caption{For the cuprate-like, insulating AFM phase labeled  $\uparrow \downarrow, \bar{z} \bar{z}$ in Fig. \ref{fig:cf_bulk}, we show: (a) the total DOS per Ni site; (b) the partial DOS projected onto the two orbitals; (c),(d) the partial DOS projected by spin component, for the two inequivalent Ni sites. Both orbital contributions are included in the spin-resolved PDOS histograms, but one can infer the orbital character of the various bands from the results in panel (b). }The vertical red line marks the Fermi energy. Histogram colors are labelled in the respective legends. There is no charge order in this phase, and the AFM order is clear from panels (c) and (d) as each site preferentially has either $\uparrow$ or $\downarrow$ occupied states below the Fermi energy. The parameters are $U/W = 0.8, J/W = 0.05$ (black star in  Fig. \ref{fig:cf_bulk}). 
  \label{fig:cf_bulk_DOS2}
\end{figure}

\subsection{The effects of reduced dimensionality}

We start by presenting in Fig. \ref{fig:cf_bulk} the phase diagram of the 2D system when both the lattice coupling and crystal-field splitting are set to zero, $\epsilon_b = \Delta_{\text{CF}} = 0$, and there is no strain. We set $t_1^{(0)} = 1, t_2^{(0)} = 0.15, t_3^{(0)} = 0.05$, so that we can meaningfully compare the 2D results with the counterpart 3D bulk results. \cite{Lee_2011,Fomichev_2020} This comparison will allow us to infer the effects of the reduced dimensionality. 

Figure \ref{fig:cf_bulk} shows a rich phase diagram. At small $U$ and  $J$, the system is a paramagnetic metal with no charge disproportionation, and with $\bar{z}$ orbitals preferentially occupied (phase labelled $00, \bar{z}\bar{z}$ in the figure legend). The degree of $\bar{z}$-polarization is indicated by the dashed contour lines. It is relatively low for larger $J$ values but increases as $J$ decreases.
Such a paramagnetic metallic state without magnetic or charge order is expected when the correlation parameters $U,J$ are small. The orbital polarization is a direct consequence of the reduced dimensionality, which favors hopping amongst the in-plane, $\bar{z}$ orbitals. It decreases with increasing $J$ because Hund's coupling favors occupation of both orbitals (when $U$ is small) to maximize the spin. We note that the ``red square'' features that appear on the diagonal in the upper-right corner of Fig. \ref{fig:cf_bulk}, and in several subsequent figures, are merely artefacts of the contour plotting algorithm. There are a few points along that diagonal which have a value of charge disproportionation very different from their immediate neighbours (it is near zero). This could be due to imperfect convergence near a magnetic phase boundary (if the true ground state was not identified), or possibly there is another kind of phase that is not well-resolved by the two-site model used in this study. Because of this stark difference in charge disproportionation, the automatic contour plotting algorithm chooses to place a contour around the points, which, being one one pixel in size, appear as little red squares. These features do not affect the main results of this work, and we ignore them in the following discussion.

Figure  \ref{fig:cf_bulk_DOS1} shows the density of states (DOS) per Ni site in this metallic phase (in units of $1/t_1^{(0)}$), plotted against the energy (in units of $t_1^{(0)}$), at the point marked by a black circle in Fig. \ref{fig:cf_bulk}. Panel (a) presents the total DOS and confirms that this is indeed a metal. Panel (b) shows the partial DOS projected onto the two orbitals. While the orbital polarization is not very large at this point, there are clearly more states with $\bar{z}$ than with $z$  character below the Fermi energy (a clear orbital band asymmetry driven by the 2D confinement). There is neither magnetic nor charge order in this phase.

\begin{figure}[t]
  \includegraphics[width=0.85\linewidth]{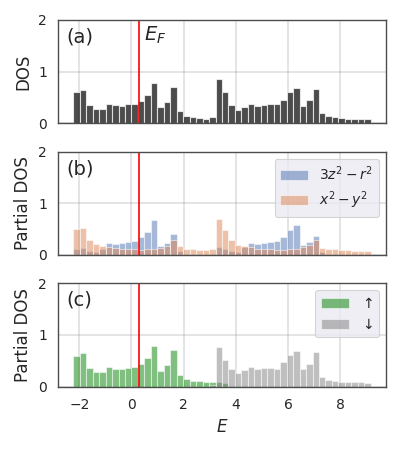}
  \caption{For the half-metal phase  $\uparrow \uparrow, \bar{z} \bar{z}$: (a) Total DOS per Ni site, (b) partial DOS with orbital projections, (c) partial DOS with spin projections. Parameters are $U/W = 0.8, J/W = 0.1$ (black square in  Fig. \ref{fig:cf_bulk}.) The orbital polarization already seen in the metallic phase in Fig. \ref{fig:cf_bulk_DOS1} is joined by FM polarization, as seen in panel (c).}
  \label{fig:cf_bulk_DOS3}
\end{figure}

\begin{figure}[t]
  \includegraphics[width=0.85\linewidth]{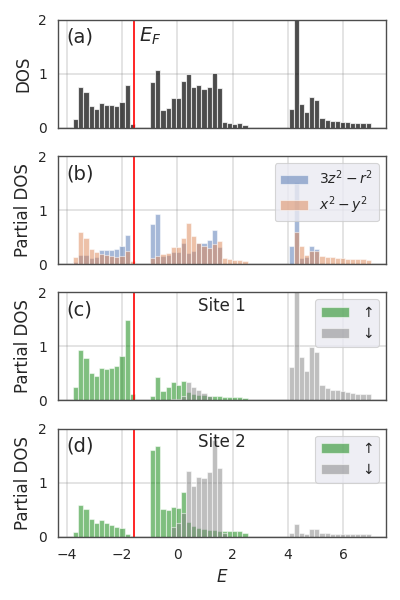}
  \caption{(a) Total DOS per Ni site (averaged over two inequivalent sites) in the charge-disproportionated insulator $\uparrow \uparrow, $ CD. Parameters are $U/W = 0.5, J/W = 0.2$ (black diamond in  Fig. \ref{fig:cf_bulk}). (b) Site-averaged orbital DOS, and (c),(d) spin-projected DOS for two inequivalent sites. Note that due to charge disproportionation, at site 1 the portion of the DOS below the Fermi level is larger, as electron density is shifted from site 2 to 1 for $\delta > 0$. }
  \label{fig:cf_bulk_DOS4}
\end{figure}

Returning to the phase diagram of Fig. \ref{fig:cf_bulk}, at large $U/W > 0.5$ and small $J$ we find an antiferromagnetic (AFM) insulator, with very strong  $\bar{z}$ orbital polarization (phase labelled $\uparrow \downarrow, \bar{z} \bar{z}$). These characteristics are verified in Fig. \ref{fig:cf_bulk_DOS2} which shows the DOS with a gap at the Fermi energy in panel (a), the partial DOS per Ni site projected on the two orbitals in panel (b), demonstrating nearly 100\% $\bar{z}$ orbital polarization (second panel), and the AFM spin ordering across the two inequivalent Ni sites in panels (c) and (d). There is no charge order in this phase. These particular results are for the point indicated by a black star in Figure \ref{fig:cf_bulk}.

This is the cuprate-like phase that we are primarily interested in. It can be thought of as the Mott insulator that emerges from the metallic phase when $U$ is increased, given that the reduced dimensionality and the low $J$ essentially turn this into a half-filled one-band problem. Its appearance in the 2D nickelate phase diagram even in the absence of favorable crystal field splitting and/or strain is one of the main results of this work. It is important to note that the Hartree-Fock approximation underestimates the effect of electronic correlations, when compared with more accurate techniques, for instance the Gutzwiller approximation \cite{Chaloupka_2008}. Thus, the cuprate-like phase can be expected to be even more robust than is shown here.

For these larger $U$ values, further increasing $J$ favours the intra-orbital spin ``aligning'' tendency  and results in the ferromagnetic (FM) half-metal phase labelled $\uparrow \uparrow, \bar{z}\bar{z} $, stable at moderate $J$. Figure \ref{fig:cf_bulk_DOS3} shows its metal-like DOS in panel (a), a small degree of orbital polarization in panel (b), and complete FM polarization in panel (c). There is no charge order in this phase. These particular results are for the point indicated by a black square in Figure \ref{fig:cf_bulk}.

\begin{figure*}[t]
\includegraphics[width=\linewidth]{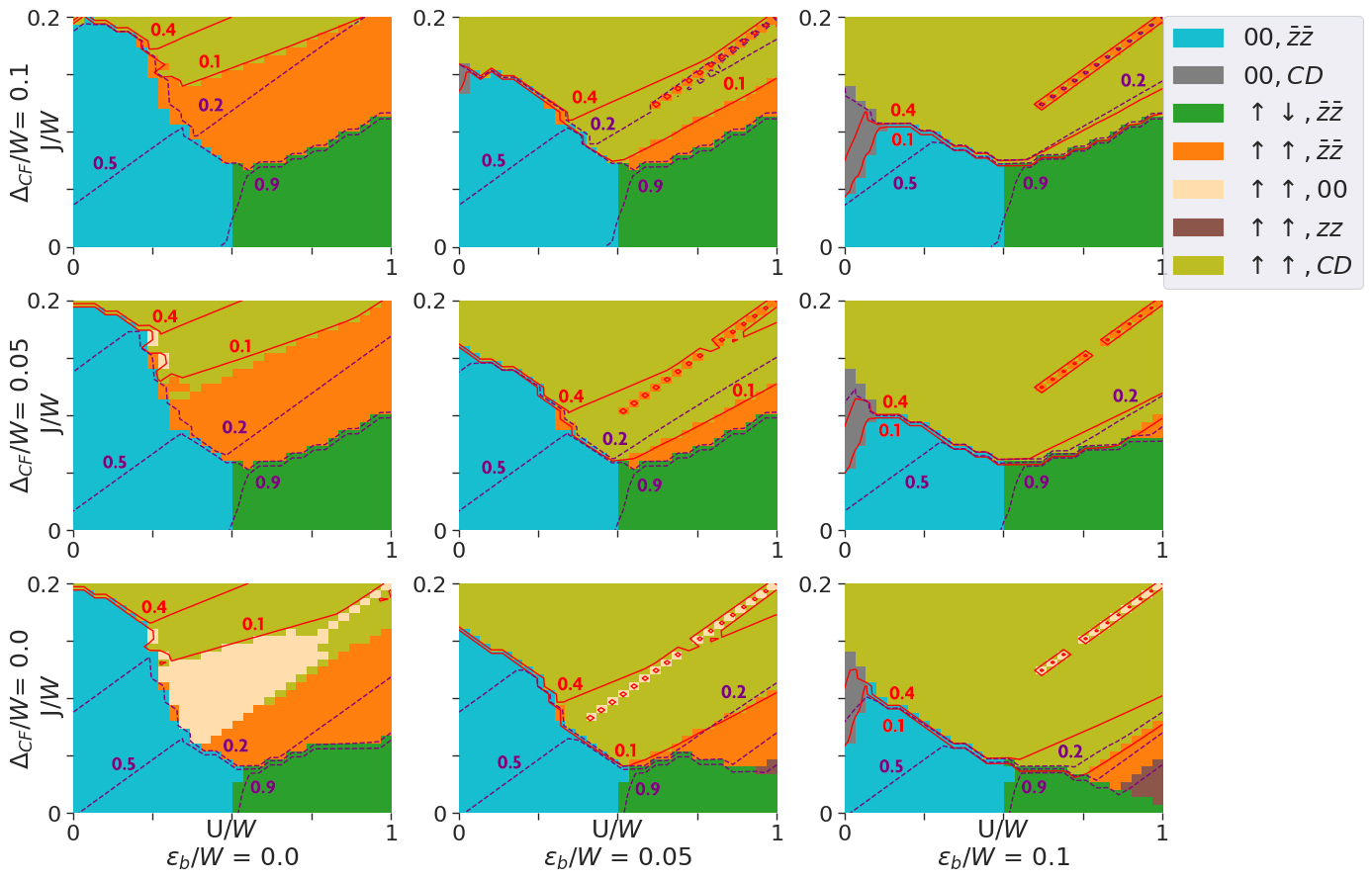}
\caption{Combined effect on the $T=0$ 2D phase diagram of the crystal field splitting $\Delta_{\text{CF}}/W$ [increasing vertically from 0 (bottom row) to 0.05 (middle row) and 0.1 (top row)], and of the electron-lattice coupling $\epsilon_b/W$ [increasing horizontally from 0 (left column) to 0.05 (middle column) to 0.1 (right column)]. Solid contours and corresponding red numbers indicate the magnitude of charge disproportionation; dashed contours with purple numbers indicate the magnitude of orbital polarization. The cuprate-like phase with nearly complete orbital order is favored by increasing $\Delta_{\text{CF}}$  and suppressed by increasing $\epsilon_b$. }
\label{fig:combo}
\end{figure*}

\begin{figure}[b!]
  \includegraphics[width=0.8\linewidth]{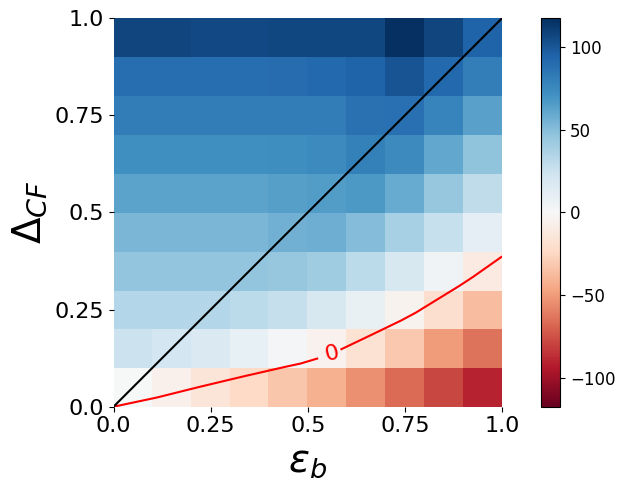}
  \caption{Change (in $\%$) of the area occupied by the cuprate-like phase in the $U$-$J$ diagram for the specified $(\Delta_{\text{CF}},\epsilon_b)$ values, {\em vs.} that of Fig. \ref{fig:cf_bulk} (i.e. for $\Delta_{\text{CF}}=\epsilon_b=0$). See text for more details. The red contour marks an unchanged area, and the black diagonal shows the line of equal strengths $\Delta_{\text{CF}} = \epsilon_b$. }
  \label{fig:meta}
\end{figure}

Approaching large $J$ values in the phase diagram favours the appearance of  finite charge disproportionation $\delta \ne 0$ (the solid red lines show $\delta= 0, 0.1, 0.3, 0.5$ contours), in order to maximize the on-site spin at one of the two sites of the unit cell.  This phase is labelled $\uparrow \uparrow$, CD in the phase diagram. This is confirmed by the results of Fig. \ref{fig:cf_bulk_DOS4}, which correspond to the black diamond shown in  Fig. \ref{fig:cf_bulk}. Panel (a) of the figure confirms that this is an insulator. Panel (b) shows that the reduced dimensionality continues to be reflected in the different partial DOS for the two orbitals, although there is only a very small $\bar{z}$ orbital polarization after summing over all occupied states. Panels (c) and (d) verify that this is an FM state with unequal charge density on the two sites of the unit cell, reflecting the CD $\delta \ne 0$. In the limit $J\rightarrow \infty$, this evolves smoothly towards full CD $\delta=1$, with spins $S=1$ and $S=0$ for the doubly occupied and for the empty site of the unit cell, respectively. In this limit, the magnetic order would be more properly labelled as $\uparrow 0$.

Finally, close to the diagonal $U=5J$, the phase diagram hosts a phase labelled $\uparrow\uparrow,00$. This is an FM with very little CD and very little orbital polarization. As further results will show, this phase becomes unstable when other ingredients are added to the model, so its appearance in a realistic system is rather unlikely.

Before moving on, it is worth comparing this 2D phase diagram to its 3D counterpart presented in Ref. \onlinecite{Fomichev_2020}. The key difference is the strong orbital polarization in 2D for small-to-moderate $J$, because the reduced dimensionality favours the occupancy of the in-plane orbital $d_{x^2-y^2}$. In particular, this allows for the appearance of the cuprate-like phase in the 2D phase diagram. By contrast, in 3D the hopping between the $z$ and $\bar{z}$ orbitals is symmetric, thus no orbital polarization is expected -- and none was found. 

Higher $J$ favors CD and reduces the orbital polarization in 2D, thus one expects more similarities in this region of the phase diagram.  At first sight this appears to be false, because in 3D the large $J$ region with CD adopts the $\uparrow0\downarrow0$ magnetic order. The difference in magnetic order is because the two-site unit cell of the 2D calculation cannot reproduce such 4-site an order, so instead it ``does the best it can'' by approximating it with the 2-site  $\Uparrow\uparrow$ ``slice'' ($\Uparrow$ means larger spin magnitude due to partial CD). We expect that using a 4-site unit cell in 2D would restore here the $\uparrow0\downarrow0$ magnetic order like in 3D. However, the larger unit cell would not affect the cuprate-like phase that is our main focus, so we continue with the two-site unit cell.

\subsection{The effect of crystal field splitting and of electron-lattice coupling }

\begin{figure*}
\includegraphics[width=\linewidth]{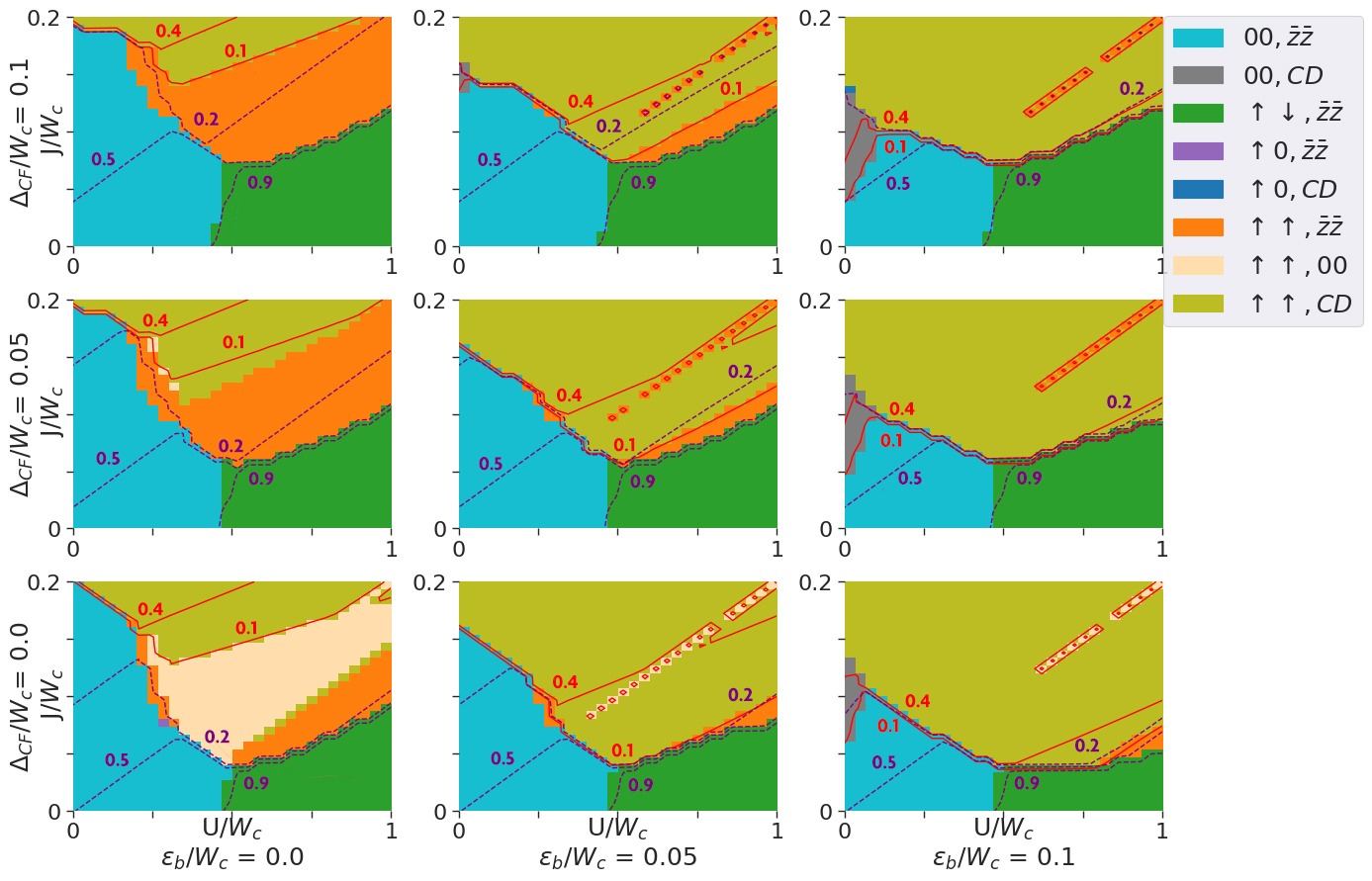}
\caption{Combined effect on the $T=0$ 2D phase diagram  of the crystal field splitting $\Delta_{\text{CF}}$ and electron-lattice coupling $\epsilon_b$ when a very large tensile strain $\epsilon = +1$ is applied.  Note that here the non-interacting bandwidth $W_t$ is roughly $e$ times smaller than in Fig. \ref{fig:combo}, according to Eq. (\ref{eq:strain}). Contours are as in Fig. \ref{fig:combo}.}
\label{fig:tensile}
\end{figure*}

\begin{figure*}
\includegraphics[width=\linewidth]{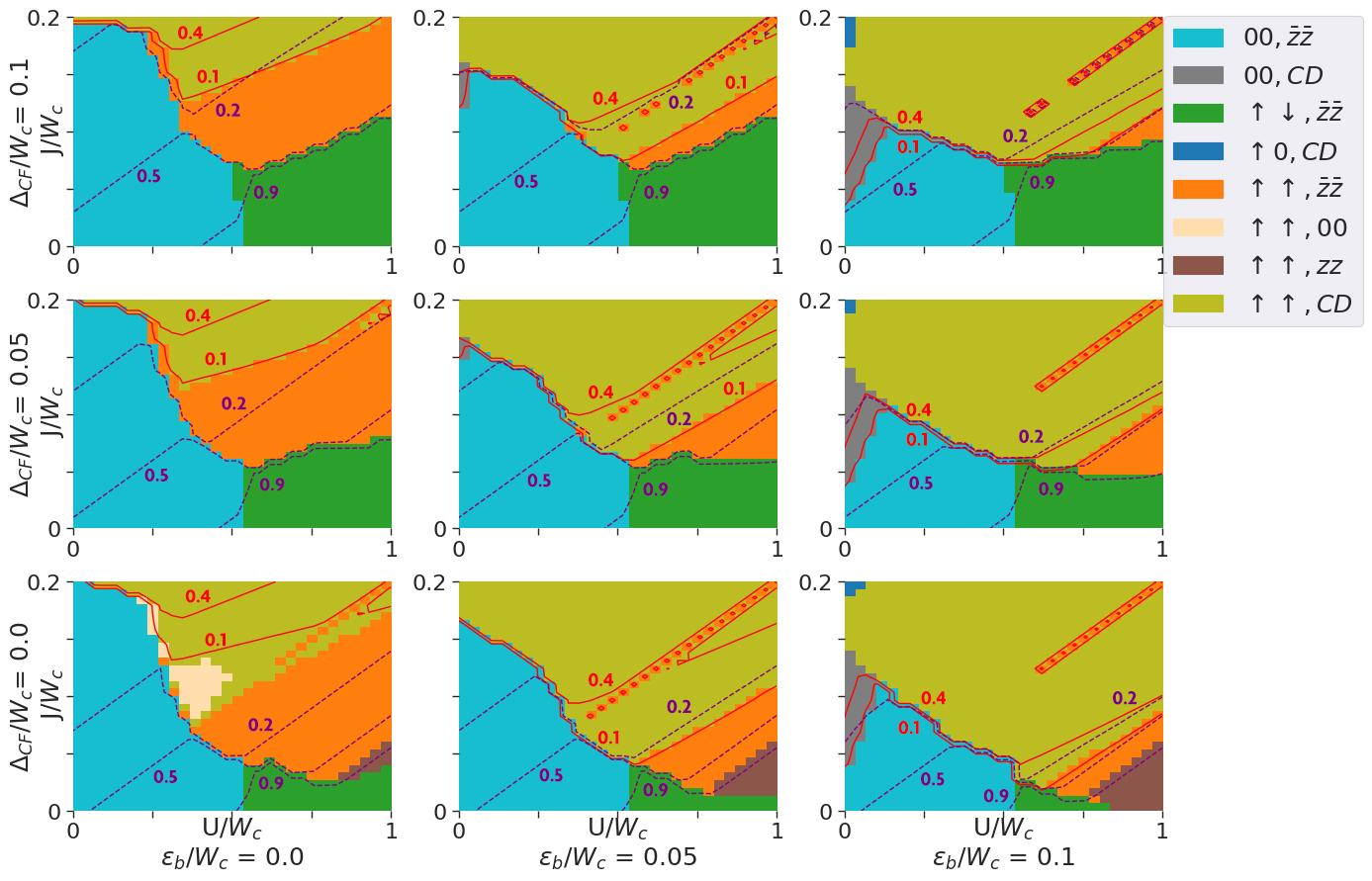}
\caption{Combined effect on the $T=0$ 2D phase diagram  of the crystal field splitting $\Delta_{\text{CF}}$ and electron-lattice coupling $\epsilon_b$ when a very large compressive strain $\epsilon = -1$ is applied.  Note that here the non-interacting bandwidth $W_s$ is roughly $e$ times larger than in Fig. \ref{fig:combo}, according to Eq. (\ref{eq:strain}). Contours are as in Fig. \ref{fig:combo}.}
\label{fig:compressive}
\end{figure*}

Figure \ref{fig:cf_bulk} confirmed the existence of a cuprate-like phase for the 2D monolayer,  even in the absence of favorable crystal field splitting $\Delta_{\text{CF}}$, or of coupling to the lattice $\epsilon_b$. We now examine the effects of these parameters on the phase diagram.  The results are shown in Fig. \ref{fig:combo}, with $\Delta_{\text{CF}}/W$ increasing vertically from 0 (bottom row) to 0.1 (top row); while $\epsilon_b/W$ increases along the horizontal from 0 (left column), to 0.1 (right column). 

As expected, adding even a moderate crystal field splitting $\Delta_{\text{CF}}>0$ leads to a substantial shift of the phase boundaries in favour of those with planar orbital ($d_{x^2-y^2} \equiv \bar{z}$) order, resulting in orbital polarization across a wider region. In particular, the cuprate-like phase expands its area significantly with increasing $\Delta_{\text{CF}}$. 

On the other hand, turning on the coupling to the lattice promotes the breathing-mode distortion that favors CD to the detriment of phases with orbital polarization. In particular, the area occupied by the cuprate-like phase is reduced, being partially pushed down by the $\uparrow\uparrow,\bar{z}\bar{z}$ phase, and partially replaced by  a new $\uparrow\uparrow,zz$ phase which has the opposite orbital polarization. The paramagnetic metal is also partially replaced by a paramagnetic CD phase at vanishing $U$ (labelled $00,$CD).

Clearly, $\Delta_{\text{CF}}$ and $\epsilon_b$ have opposing effects on the stability of the cuprate-like phase.

To illustrate this in a more quantitative way, we generate a \textit{meta} phase diagram, shown in Fig. \ref{fig:meta}. For each pair $(\Delta_{\text{CF}}, \epsilon_b)$ shown in Fig. \ref{fig:meta}, we generate the corresponding $U$-$J$ phase diagram, and compute the area $A_{c}$ occupied by the cuprate-like phase. We compare this to the area  $A$ it occupies when $\epsilon_b = \Delta_{\text{CF}} = 0$. The quantity $r_c = (A_c / A - 1)\times 100\%$, giving the percentage area increase with respect to the $\Delta_{\text{CF}} = \epsilon_b = 0$ point, is color-plotted in Fig. \ref{fig:meta}. It confirms that when $\Delta_{\text{CF}}=\epsilon_b$,  the crystal field splitting increases the area occupied by the cuprate-like phase more efficiently than electron-lattice coupling suppresses it. However, we note that the ratio of the actual energy contributions to the ground-state energy coming from these two terms can be a rather complicated and not necessarily monotonic function of $\Delta_{\text{CF}}/\epsilon_b$, given that the orbital polarization and the lattice distortion, respectively, are also involved in their expressions. However, these variations are limited to factors of at most $\frac{1}{2}$ and would not change out fundamental conclusion.

\subsection{The effect of strain}

Thin film nickelates and heterostructures  are grown on a variety of substrates, leading to lattice mismatch. The resulting biaxial strain state affects the electronic and orbital properties, as numerous experiments have demonstrated (see discussion in Sec. \ref{sec:intro}). Strain  is expected to be an important  control variable for nickelate monolayers; we now analyze its effect.

According to the SIOP model (see Sec. \ref{sec:intro}), strain controls the energy balance between the in-plane and the out-of-plane orbitals in the Ni octahedra, favoring $\bar{z}$-polarization for tensile strain and  $z$-polarization for compressive strain. In this view, the strain's impact on the phase diagram is analogous to that of a crystal field splitting, which has already been discussed. 

However, this ignores another important factor: applying compressive (tensile) strain also changes the relative magnitude of the hopping integrals $t_1, t_2, t_3$. As found in the 3D bulk study \cite{Fomichev_2020}, the ratio $t_2 / t_1$ is very important in determining the physics in the moderate $U$, moderate $J$ region of the phase diagram -- for two reasons. 

One is that bipartite lattices (i.e ones with pure $t_1$ hopping, $t_2 = t_3 = 0$) have the strongest Fermi surface nesting tendency and are thus susceptible to $\pi(1,1)$ order \textit{of some kind}. On the other hand, a robust value of $t_2$ has been shown to benefit FM states due to a Stoner-like effect wherein a concentration of density near the Fermi surface benefits FM ordering. This is because  the Stoner criterion is $D(E_F) U \sim 1$, where $D(E_F)$ is the density of states at the Fermi energy, so a strong enhancement of the density of states (which is exactly what a relatively larger $t_2$ is found to promote) reduces the interaction strength at which FM ordering is expected. These two mechanisms lead to an intense competition in the moderate $U$ moderate $J$ sector of the phase diagram between various FM and AFM phases.

This argument, however, assumes that the other parameters (apart from $\Delta_{\text{CF}}$, see above) are not significantly affected by the strain. This is not true because hopping integrals also influence the wavefunctions of the ``molecular orbitals'' represented by our effective model, and thus the effective $U,J,\epsilon_b$ values. Without detailed modeling that goes well beyond the scope of this work, it is hard to know how  $U/W, J/W, \epsilon_b/W$ evolve as strain changes the bandwidth $W$ of the free electron band. We therefore analyze the effect of the strain on the phase diagram by varying the hopping integrals with the strain, as defined in Eq. (\ref{eq:strain}), while the ratios of the other parameters cover the same ranges as before
({\it eg.}, $U/W\in [0,1]$ for each strain-modified value of $W$).

We begin by analyzing the effect of tensile strain on the $U$-$J$ phase diagram, at various strengths of electron-lattice coupling and crystal field splittings. We use a very large strain value $\epsilon=1$ for illustrative purposes. The results are shown in Fig. \ref{fig:tensile}, with panels organized like in Fig. \ref{fig:combo}. Note that the non-interacting bandwidth, which we call $W_c$, is now roughly $e\approx 2.7$ times smaller than its value $W$ in the absence of tensile strain.

The main change in the $\Delta_{\text{CF}} = \epsilon_b = 0$ phase diagram is the expansion of the pure FM phase at the expense of the $\bar{z}$ polarized FM phase (compare bottom leftmost panels in Figs. \ref{fig:tensile} and \ref{fig:combo}); the other phases are roughly unaffected. At first one may expect that no change should occur, given that we scaled $U,J$ with the new bandwidth $W_c$. However, as just mentioned, the ratios between $t_1$, $t_2$ and $t_3$ hoppings change significantly, see Eq. (\ref{eq:strain}), revealing that it is such details of the bandstructure which decide which phases are stable in the central region of the phase diagram. The cuprate-like phase of interest to us is little affected, proving that it is robust against such variations. The phase diagrams corresponding to finite $\Delta_{\text{CF}}$ and/or $\epsilon_b$ are fairly similar to their strain-free counterparts. 

Overall, we find that the $\bar{z}$-polarized phases are more stable under tensile stress, and in particular the cuprate-like phase increases its area in all cases studied. This occurs largely at the expense of the middle FM phases due to the competition mentioned earlier. 

The corresponding results in the presence of a very significant compressive strain $\epsilon=-1$ are shown in Fig. \ref{fig:compressive}. The cuprate-like phase is more suppressed here than in the absence of strain, although it still survives even in the least favorable circumstances of large $\epsilon_b$ and vanishing $\Delta_{\text{CF}}$. In exchange, the FM $z$-polarized phase becomes more robust and appears even in the $\Delta_{\text{CF}} = \epsilon_b = 0$ phase diagram. This ties is with experimental observations and the SIOP model which find preferential $z$ orbital occupation under compressive strain. For completeness, we mention the appearance of a phase labelled $\uparrow0,$CD, which is the $\delta\rightarrow 1$ limit of the $\uparrow\uparrow,$CD phase, as discussed before. Despite the smooth evolution between the two, we mark them with  different colors to indicate where full CD has been reached.

These results show that so long as all the other parameters scale with the bandwidth and $J, \epsilon_b$ are not so large as to drive CD, the effect of strain coming from variations of the relative strengths of various hopping integrals are so as to favor $\bar{z}$-polarization for tensile strain, and $z$-polarization for compressive strain. This is roughly consistent with the SIOP model, although clearly the situation is much more complex than predicted by the latter due to details of the density of states. Our results show the appearance of a $z$-polarized phase even for strong tensile strain if the coupling to the lattice is strong enough; we also expect that the $\bar{z}$-polarized cuprate-like phase will survive on the $\Delta_{\text{CF}}<0$ side of the phase diagram with compressive strain applied, if $\epsilon_b$ is not too large and $\Delta_{\text{CF}}$ is not too negative. Neither of these results are predicted by the SIOP model, but they may begin to qualitatively explain the unusual relationship between strain and degree of orbital polarization found in some thin films (mentioned in the Introduction).

\section{Conclusions}\label{sec:concl}

We used an effective two-band Hubbard-Kanamori model coupled to lattice distortions,   to investigate the complex interplay of orbital, charge, spin and lattice behaviour in a 2D monolayer of rare-earth nickelates, with a focus on understanding whether/when a cuprate-like phase becomes stable\cite{Chaloupka_2008}.

We find that the reduced dimensionality indeed favors such a phase, which becomes more robust in the presence of tensile strain and/or positive crystal field splitting $\Delta_{\text{CF}}>0$. The cuprate-like phase would likely be further enhanced if the electron correlations were treated more accurately by going beyond the Hartree-Fock approximation, because correlations tend to favour insulating AFM behaviour, as is often seen with the orbitally selective Mott transition. 

On the other hand, strong coupling to the lattice (favoring lattice distortions and charge disproportionation) and/or compressive strain (favoring $z$-orbital polarization) are detrimental to the stability of this phase. Of course, so is a negative crystal field splitting.

Crucially, the 2D phase diagram
is qualitatively different from its 3D counterpart \cite{Fomichev_2020} in the region where this cuprate-like phase appears. This is reminiscent of the recent experimental observation that a 2D monolayer has a ground-state different from that found in thin films or in bulk, \cite{Disa_2017} although that monolayer is apparently on the $\Delta_{\text{CF}}<0$ side, which is less favourable to a cuprate-like ground-state. 

To conclude, this work offers further support to the idea that a cuprate-like phase might be stabilized in a 2D monolayer in favorable conditions, as well as some guidance as to what those conditions are. We therefore hope that it will stimulate both further, more detailed theoretical work to identify the specific heterostructures that host this cuprate-like phase, and experimental work searching for it.

\begin{acknowledgments}
We would like  to  acknowledge the invaluable help of Evgenia Krichanovskaya with the  graphic  design  of  the  various  plots  and  diagrams. This work was supported by the UBC Stewart Blusson Quantum Matter Institute, the UBC SBQMI Quantum Pathways program, the Max-Planck-UBC-UTokyo Center for Quantum Materials and the Natural Sciences and Engineering Research Council of Canada.
This research was supported in part through computational resources and services provided by 
Advanced Research Computing at the University of British Columbia.
\end{acknowledgments}

\appendix

\section{\label{app:Hopping}Hopping matrix elements}

In this appendix we provide the expressions for the Fourier-transformed hopping matrix elements $t_{ab}(\mathbf{k})$ for first, second and fourth nearest neighbour hopping in 2D (cf. Ref. \onlinecite{Fomichev_2019} for bulk nickelates hopping). They are given by
\begin{widetext}
\begin{equation*}
    t_{zz}(\textbf{k}) = - \frac{t_1}{2}[\cos(k_x a) + \cos(k_y a)] - \frac{t_3}{2}[\cos(2k_x a) + \cos(2k_y a)] - 2t_2\cos(k_x a)\cos(k_y a)
\end{equation*}
\begin{equation*}
    t_{\bar{z}\bar{z}}(\textbf{k}) = - \frac{3t_1}{2}[\cos(k_x a) + \cos(k_y a)] - \frac{3t_3}{2}[\cos(2k_x a) + \cos(2k_y a)] + 6t_2\cos(k_x a)\cos(k_y a)
\end{equation*}
\begin{equation*}
    t_{z\bar{z}}(\textbf{k}) = \frac{\sqrt{3}t_1}{2}[\cos(k_x a) - \cos(k_y a)] + \frac{\sqrt{3} t_3}{2}[\cos(2k_x a) - \cos(2k_y a)]
\end{equation*}

\end{widetext}

\section{Numerical Methods}\label{sec:numerics}
Once equipped with our Hamiltonian matrix, self-consistency equations and mean field parameters, we are ready to solve for the model's ground state. As in Ref. \onlinecite{Fomichev_2020}, we start by guessing an initial set of mean field parameters, compute the eigenvectors of the Hamiltonian at all lattice points, evaluate the mean-field parameters from the self-consistency equations using all occupied energies, and compare the new mean field parameters to the initial ones. We then iterate updating the guesses until self-consistency is attained.

\begin{figure}
  \includegraphics[width=\linewidth]{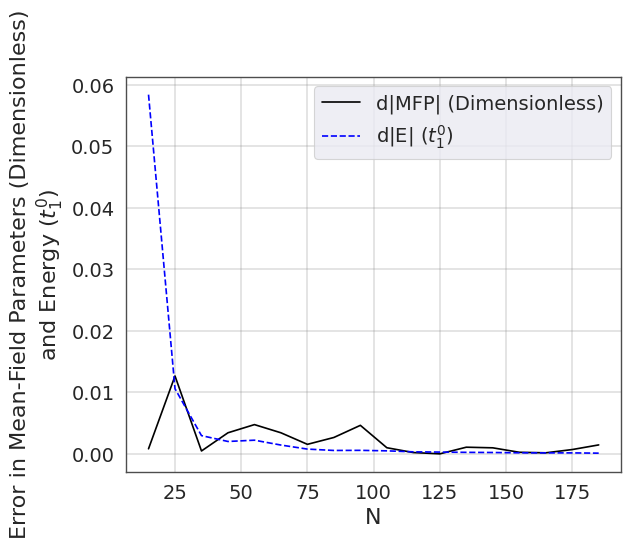}
  \caption{Euclidean distance between mean field solutions (solid black line) and difference in ground state energies per site (dashed blue line) shows convergence as a function of $N$, the number of momentum points per dimension. The hopping parameters are $t_1^{(0)} = 1, t_2^{(0)} = 0.15, t_3^{(0)} = 0.05$, and the remaining model parameters were chosen to be $U=J=\Delta_{CF}=\epsilon_b=0$.}
  \label{fig:momentum convergence}
\end{figure}

\begin{figure}
  \includegraphics[width=\linewidth]{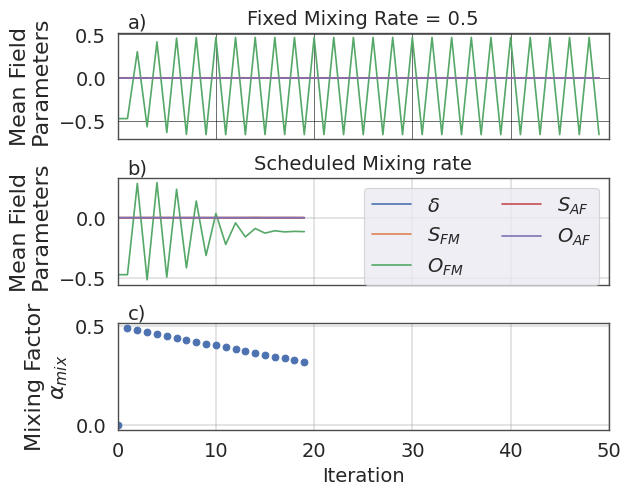}
  \caption{Comparing the mean-field parameters during the iterative process obtained from (a) using a fixed mixing rate $\alpha_{mix}=0.5$, vs. (b) using the scheduled mixing rate. The iteration dependence of the scheduled mixing rate used in panel (b) is shown in panel (c). As can be seen from panel (a), using a fixed mixing rate results in the iteration ``cycling'' between the same few points, until the iteration sequence is aborted upon reaching the maximum iteration limit. By comparison, panel (b) shows that the scheduled mixing rate at the same value of parameters leads to rapid convergence in merely 15 iterations. Model parameters are the same as in Fig. \ref{fig:momentum convergence}, except $J = 1.2$. The number of momentum points per dimension $N$ is set to 100.}
  \label{fig:FixedVsScheduled}
\end{figure}

In all calculations, we compute and store all matrix variables that are independent of the mean field parameters, such as the hopping terms, to avoid repeating calculations over iterations. In every diagonalization, we then access the computed values by their momentum index.

The next approximation is to select a finite grid of momentum points to approximate the crystal's Brillouin zone. We let $N$ = $N_x$ = $N_y$ be the number of momentum points in each dimension such that there are $N^{2}$ momentum points. To choose an appropriate $N$, we perform a preliminary study and choose the smallest $N$ such that the mean field solution and ground state energy converge up to our desired limit, of $10^{-3}$. Fig. \ref{fig:momentum convergence} shows the change in ground state energy per site $d|E| = |E_{HF}^{(n)} - E_{HF}^{(n-1)}| / N$ and the residual $\epsilon^{(n)} = | \mathbf{w}^{(n)} - \mathbf{v}^{(n)} |$ as a function of $N$ (here $n$ is the final iteration upon which convergence has been achieved, and $\mathbf{w}, \mathbf{v}$ are defined in Sec. \ref{sec:calc}).

After finding the occupied energies and the new mean field parameters, we now have to choose the new mean field ansatz for the next iteration. The standard approach is to take the previous answer as the new input. Otherwise, as done in Ref. \onlinecite{Fomichev_2020} or Ref. \onlinecite{Banerjee_2016}, one can take the new ansatz to be some linear combination of previous iteration steps, with so-called mixing rates $\alpha_{mix}^{(i)}$ controlling the relative contributions from previous steps.

Instead, we suggest a new method, inspired by techniques in the field of machine learning optimization, where it is common to control learning steps by a scheduled learning rate. In this field, to ensure convergence of model training, it is common to ``schedule'' the learning rate to approach zero, taking smaller steps as the iterative sequence goes on. 

We apply this idea of scheduling to the mixing rates $\alpha_{mix}$ of the Hartree-Fock iteration, defining the next iteration's original guess $\mathbf{w}^{(n)}$ to be
\begin{equation}
    \mathbf{w}^{(n)} = [1 - \alpha_{mix}(n)] \mathbf{v}^{(n-1)} + \alpha_{mix}(n) \mathbf{w}^{(n-1)},
\end{equation}
where $\mathbf{w}^{(n)}$ is the initial guess for the mean field parameter vector for iteration $n$, $\alpha_{mix}(n)$ is the mixing rate at iteration $n$, and $\mathbf{v}^{(n-1)}$ is the mean field parameter vector obtained from the guess $\mathbf{w}_{(n-1)}$. The choice $\alpha_{mix}(n) = 0$ would correspond to the simple rule $\mathbf{w}^{(n)} = \mathbf{v}^{(n-1)}$ referenced in Sec. \ref{sec:calc}. We choose the scheduled mixing rate $\alpha_{mix}(n)$ such that it is bounded by $0.5$, and asymptotically approaches zero. In this study we use a sigmoid function $\alpha_{mix}(n) = 1/(1 + \exp(-n\beta/T))$ for the mixing rate, where $n$ is the current iteration, $\beta = 3$, and $T = 250$ is the maximum iteration limit. This choice was found to yield fast convergence for most parameter values.

With this choice for the Hartree-Fock iterations, we found that convergence was reached with many fewer iterations. Moreover, many iteration sequences that would not converge with the fixed mixing rate could now easily converge with the scheduled mixing rate. One example of this improvement is shown in Fig. \ref{fig:FixedVsScheduled}: in panel (a) a fixed mixing rate $\alpha_{mix} = 0.5$ is used, and no convergence is achieved, as the mean field parameters keep oscillating between two fixed points. Meanwhile, in panel (b) a scheduled mixing rate is used (its dependence on the iteration is shown in panel (c)): convergence to the ground state is achieved within 15 iterations. Overall, with this method we find that convergence improves from about 80\% to 99.9\% across the studied phase diagrams within the chosen maximum iteration limit of 150.

\begin{figure}
  \includegraphics[width=\linewidth]{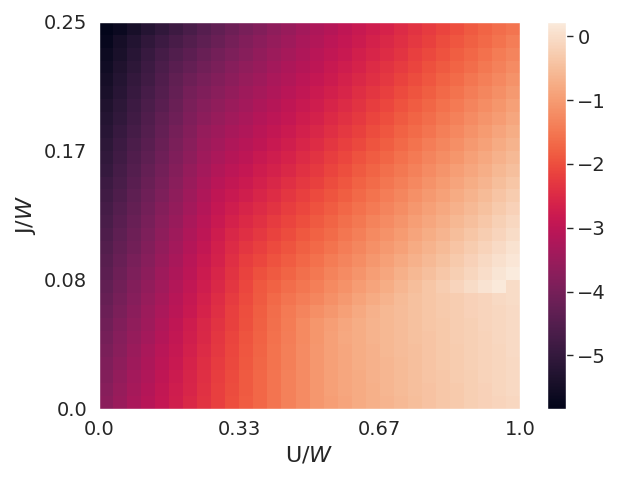}
  \caption{Ground state energy of the non-interacting model, with the same model parameters as in Fig. \ref{fig:momentum convergence}. The number of momentum points per dimension $N$ is set to 120.}
  \label{fig:EnergyDiagram}
\end{figure}

With this procedure, we first perform an exhaustive search. Using equally spaced points in the mean field parameter space, we find that the number of different solutions is much smaller than the numbers of guesses. We then use those found states as the only guesses and confirm they yield the same results as the exhaustive search. We then restrict our search to these unique vectors in further calculations for time efficiency, by avoiding redundant solutions. Thus equipped with a robust iteration procedure, we then compute the iteration across all points in the concerned phase diagrams.

By the nature of Hartree-Fock calculations, one cannot ever be sure that all consistent solutions have been found. We gain confidence in our solution's validity by noting that the identified ground state energy is a smooth function of the studied phase diagram axes, as shown in Fig. \ref{fig:EnergyDiagram}. If some of these states were metastable, that would show as a discontinuous jump their energy.

At this stage, calculations can be parallelized across four levels, namely momentum points, input ansatz, phase diagram points, and even entire phase diagrams. We vectorize calculations over all momenta, distribute phase diagram points across multiple processes, try different ansatzes serially (no parallelization), and parallelize complete phase diagrams such as Fig. \ref{fig:cf_bulk} across computing nodes. All our calculations were carried out on the University of British Columbia's Advanced Research Computing cluster Sockeye. To produce a diagram as in Fig. \ref{fig:cf_bulk}, we use a typical Sockeye node with 32 Dell EMC R440 CPU cores and 12GB of RAM for around 20 minutes. Given enough compute nodes, computing the 100 phase diagrams (90,000 individual phase diagram points, each with 5-10 starting ansatzes, 10-100 iterations per ansatz, and 14,400 16x16 matrix diagonalizations per iteration) required to produce the meta phase diagram in Fig. \ref{fig:meta} took under 4 hours, whereas without any parallelization this would require on the order of 1,000 hours. All code used to produce the results for this paper can be found on
GitHub.\footnote{\url{https://github.com/chavezrodz/Hartree-Fock-Solver/tree/main}}

\bibliography{apssamp}
\bibliographystyle{apsrev4-2}

\end{document}